# The first digestive movements in the embryo are mediated by mechanosensitive smooth muscle calcium waves

Nicolas R. Chevalier

*Laboratoire Matière et Systèmes Complexes, Université Paris Diderot/CNRS UMR 7057, Sorbonne Paris Cité, 10 rue Alice Domon et Léonie Duquet, 75013 Paris, France*

Correspondence should be addressed to: nicolas.chevalier@univ-paris-diderot.fr

## Abstract

Peristalsis enables transport of the food bolus in the gut. Here, I show by dynamic ex-vivo intra-cellular calcium imaging on living embryonic gut transverse sections that the most primitive form of peristalsis that occurs in the embryo is the result of inter-cellular, gap-junction dependent calcium waves that propagate in the circular smooth muscle layer. I show that the embryonic gut is an intrinsically mechanosensitive organ, as the slightest externally applied mechanical stimulus triggers contractile waves. This dynamic response is an embryonic precursor of the "law of the intestine" (peristaltic reflex). I show how characteristic features of early peristalsis such as counter-propagating wave annihilation, mechanosensitivity and nucleation after wounding all result from known properties of calcium waves. I finally demonstrate that inter-cellular mechanical tension does not play a role in the propagation mechanism of gut contractile waves, unlike what has been recently shown for the embryonic heartbeat. Calcium waves are an ubiquitous dynamic signaling mechanism in biology: here I show that they are the foundation of digestive movements in the developing embryo.

## Introduction

The intestine is a peristaltic pump: smooth muscle located in the gut wall constricts and the propagation of this constriction along the gut tube pushes the food bolus forward. By actively replenishing the nutrient source required for growth, peristalsis enabled pluricellular organisms to grow beyond the size limit imposed by diffusive food transport. Peristalsis is most clearly evidenced in the gastrointestinal tract, but is also central to uterine fluid-, sperm-, urine-, and even blood-flow in some animals. Motility in the adult gut is a complex phenomenon featuring several different patterns of movement (1) and controlled by two cell networks, the enteric nervous system (ENS) and the interstitial cells of Cajal (ICC). The ENS presents a great variety of neuronal cell types (excitatory, inhibitory, mechanosensitive etc.) that communicate with the help of more than 35 different neurotransmitters (2).

In recent years, a few groups involved in the budding and exciting research field of *developmental physiology* have highlighted how peristalsis movements emerge during embryonic development. Using different animal models (chicken (3), mouse (4,5), zebrafish(6)), they have all shown that, unlike adult peristalsis, the earliest detectable motility patterns in an embryo are purely myogenic (muscular), they involve neither neurons nor ICCs. We have recently released a comprehensive description (3) of peristalsis wave propagation patterns in the chicken embryo and of their main phenomenological physical characteristics (see VideoS1): wave nucleation sites give rise to two waves traveling in opposite directions at a constant speed in the range ~10-50 µm/s; when two oppositely traveling waves meet, they annihilate; waves are calcium-dependent, as they vanish in calcium-free medium and are affected by calcium channel blockers (cobalt chloride); waves propel luminal content at least as from E10 in chick (E17.5 in mice (7)), because green bile is present in the lumen along the entire midgut as from this age. Because of its simple geometry (a tube) and because only one cell-type is involved (smooth muscle), the embryonic gut provides a tractable experimental model of how contractile waves emerge and propagate in biological systems.

## Results

**Calcium waves underlie contractions in the early embryonic gut.** I used the cell-permeant intracellular calcium concentration indicator Fluo4-AM. The different histological layers of the gut had to be exposed to allow for permeation of the indicator. To this end, I performed thin transverse embryonic chick gut slices and imaged them after dye loading on an inverted confocal microscope (Fig.S1a & Materials and Methods). The sample exhibited spontaneous, radial contractions occurring with a frequency of ~2-3/min and lasting for at least 4 h (Fig.1a and VideoS1). The transverse view clearly shows how the contraction obliterated the lumen. Contractions started at a point in the smooth muscle ring and quickly propagated circumferentially around the ring, leading to an overall radial constriction of the gut section. This radial constriction then propagated longitudinally, along the long-axis of the intestine (VideoS2). A key advantage of the transverse-slice configuration is that displacements induced by the spontaneous contraction are radial, so that cells remain roughly in the optical plane of the confocal microscope during imaging. This minimizes fluorescence fluctuations due solely to displacements of the sample in and out of the confocal imaging plane. VideoS3 shows a representative ($n=7$ samples from $n=4$ embryos, $n=14$ movies examined) time-lapse movie of calcium dynamics in a slice of living E9 jejunum. The contraction is seen to initiate at a point at 10-11 o'clock and to propagate in clockwise and counter-clockwise directions in the explant (see Fig.1b). The propagation of the contraction is accompanied by propagating flashes and flickers of fluorescence in individual cells (VideoS3 & Fig.S2). The cell layer in which flashes occurred was also the one that contracted circumferentially, indicating that smooth muscle cells were the most likely source of calcium transients. Groups of cells located at the periphery of the gut slice (Fig.1b, VideoS3) were also statically fluorescently labeled by Fluo4-AM. Post-calcium imaging immunohistochemical staining (Fig.1c) shows that the latter were mostly enteric ganglia (Tuj staining, see superposition of Fig.1b&c in VideoS4).

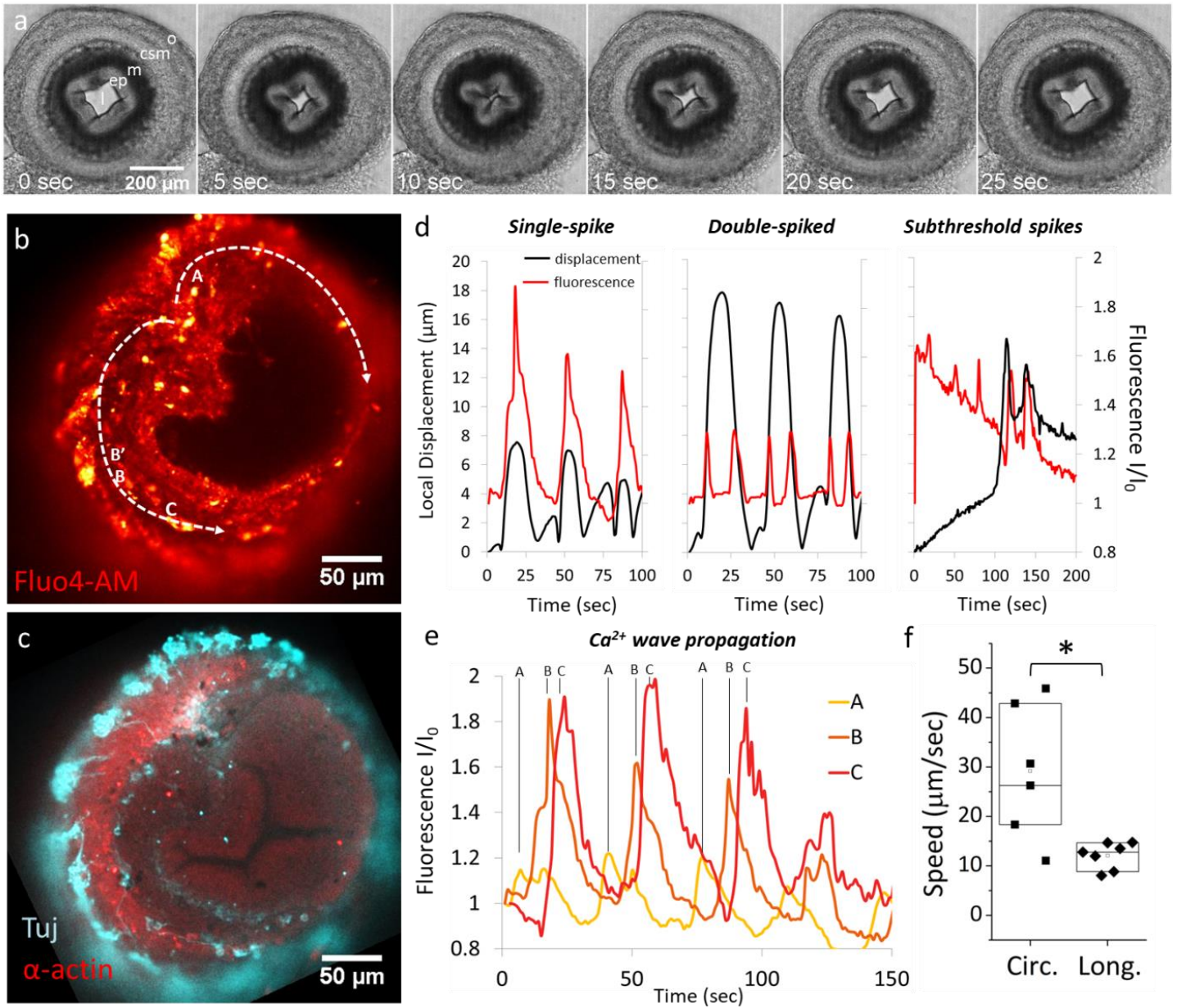

**Fig.1 : Calcium wave imaging on living embryonic gut explants.** (a) Brightfield still shots (VideoS2) of one cycle of circular smooth muscle constriction of an E9 gut transverse section embedded in agarose gel. l: lumen, e: epithelium, m: mucosa, csm: circular smooth muscle, o: outer layer (myenteric nerve plexus, not yet differentiated longitudinal smooth muscle, serosa). (b) Confocal fluorescence imaging, E9 jejunum transverse section after loading with calcium indicator. Dashed white arrows indicate the propagation direction of calcium waves in this sample, cf VideoS3. Fluorescence intensity analysis presented in d) and e) were performed on cells located at positions A, B, B' and C. (c) Immunohistochemical staining of smooth muscle (red, α-actin) and enteric neurons (Tuj, βIII-tubulin, cyan) of the slice presented in b), after calcium imaging. (d) Three representative patterns of displacement and fluorescence for different smooth muscle cells: single-spike (point B in (b) & VideoS3), double-spike (point B' in (b) & VideoS3), subthreshold spikes (from VideoS5). (e) Fluorescence as a function of time at points A, B and C in b). (f) Circumferential ($n=6$) and longitudinal ($n=7$) propagation speed of the contractile wave, E9 jejunum. Each point corresponds to a different sample. A star indicates a statistically significant difference ($p<0.05$, Mann-Whitney two-tailed test).

In contrast, the localization of the propagating calcium wave coincided with the α-actin positive circular smooth muscle ring (Fig.1b&c, VideoS4). I devised a plugin that tracks the position of a cell in the slice during the contraction, and reads out the fluorescence intensity in a small region in its immediate vicinity (Fig.1d, Materials & Methods). In the majority of cases ($n=13$ out of 22 cells analyzed from $n=3$ samples from $n=3$ embryos), the displacement peak coincided with the fluorescence increase (single-spike, Fig.1d left). In some rarer cases ($n=6/22$), two distinct calcium spikes (Fig.1d, middle) occurred for a single contraction event. I also observed subthreshold calcium spikes ($n=4/22$), i.e. fluorescence fluctuations that were not accompanied by a contraction (Fig.1d, right, arrows & VideoS5). The fluorescence intensity change $I/I_0$ of these subthreshold spikes were typically lower ($\Delta I/I_0 \sim 0.2$-$0.4$) than the ones that accompanied a contraction ($\Delta I/I_0 \sim 0.4$-$1$). Fig.1e shows the propagation of the fluorescence signal: the signal reaches its maximum intensity first at the wave nucleation point A (Fig.1b), and then at more distant points (B and C in Fig.1b). When counter-propagating calcium waves met along the gut circumference, they annihilated (VideoS6). The speed of the propagating circumferential calcium wave was equal to that of the contraction, $v_{circ} = 29.2 \pm 6.8$ μm/s (Fig.1f, $n=20$ waves, $n=6$ samples). This speed is significantly higher than the longitudinal propagation speed, $v_{long} = 12.1 \pm 1.3$ μm/s ($n=7$) measured at E9 in similar conditions (Fig.1f, data reproduced from (3)).

**Propagating contractions are mediated by gap-junctions.** Embryonic heart contraction waves have recently been shown to be independent of gap-junctions (8). To determine the role of gap junctions in the propagation of embryonic gut contractile waves, I used two pharmacological inhibitors: enoxolone (18β-Glycyrrhetinic acid) and heptanol. I placed E7 and E9 embryonic guts on Anodisc membranes on DMEM medium at 37.5°C (see Fig.S1b,c & Materials and Methods) and, after recording the native activity, added stepwise increasing concentrations of inhibitor (VideoS7). Fig.2a&b show representative motiligrams (see Materials & Methods for details); Fig.2c&d show the quantitative contractile wave frequency changes induced by enoxolone ($n=5$) and heptanol ($n=5$) in the jejunum and hindgut. Enoxolone 33 μM triggered an increase of contractile frequency by a factor ~2, followed 20-30 min after application by a gradual decrease of contractile wave amplitude (Fig.2a, VideoS7). The subsequent increase to 100 μM led to very quick fading of contraction amplitude and to the eventual cessation of wave generation and propagation (Fig.2a, VideoS7). The application of 100 μM heptanol at E7 led to a marked decrease of contractile frequency (2/5 samples), or to its cessation (3/5 samples). 500 µM heptanol completely abolished motility in all samples (5/5). The effects of enoxolone at stage E9 (*n*=5, Fig.S3) were identical to those at E7 (Fig.1a,c). Importantly, the effects of enoxolone were reversible, i.e., motility was recovered after the drug was washed away (*n*=6, Fig.S4). This shows that the abolition of gut motility at 100 µM enoxolone was not the result of drug toxicity at these concentrations, but was due to gap junction inhibition.

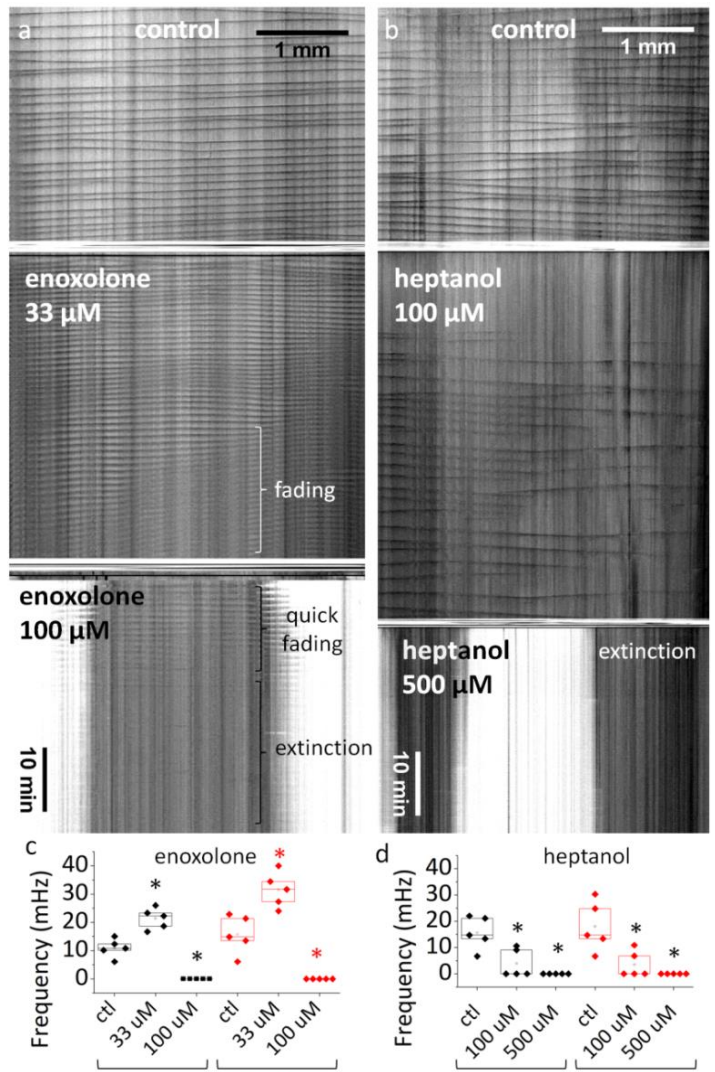

**Fig.2 : Effect of gap junction inhibitors on motility. (a) Representative motiligrams of E7 jejunum in control conditions, after addition of 33 μM and 100 μM enoxolone. (b) Motiligram of E7 jejunum in control conditions, after addition of 100 μM and 500 μM heptanol. (c) & (d) Change of contractile wave frequency in jejunum (black) and hindgut (red) in control conditions and after addition enoxolone or heptanol. A star indicates a statistically significant difference (p<0.05, Mann-Whitney two-tailed test) with respect to control conditions.**

**The early embryonic gut is mechanosensitive.** Pressure can induce contractions in the adult gut (9). To examine the response of embryonic guts to mechanical stimulation, I used a thin, pulled Pasteur pipette with a glass bead formed at the tip (Fig.3a, top). I found that gently pressing the gut with this bead consistently (>25 pinches at stage E8-E9 on *n*=6 samples) led to the nucleation of two waves traveling away from the point where pressure was applied (Fig.3a,b and VideoS8). Wave nucleation could be elicited from all gut segments (jejunum, ileum or hindgut). This reflex is extremely sensitive as even a light, delicate brush (~1-2% deformation in a ~10 µm² area) with a very thin pipette was sufficient (*n*=10/10 brushes on *n*=4 samples) to trigger wave nucleation (Video S8). I could elicit waves by mechanical stimulation at stage E6 (VideoS8, *n*=4/4 samples), the earliest stage at which physiological spontaneous propagating contractions occur (3). I further found that pre-incubating embryonic guts for 30 min with tetrodotoxin (TTX, 1 µM), a potent neuronal inhibitor (Na$^+$ channel blocker), did not alter the response of the guts to mechanical stimulation (Fig.3c & VideoS8, E9 *n*=5/5 samples). We have previously shown that spontaneous contractile wave activity is TTX-insensitive at stages E7 and E9 (3).

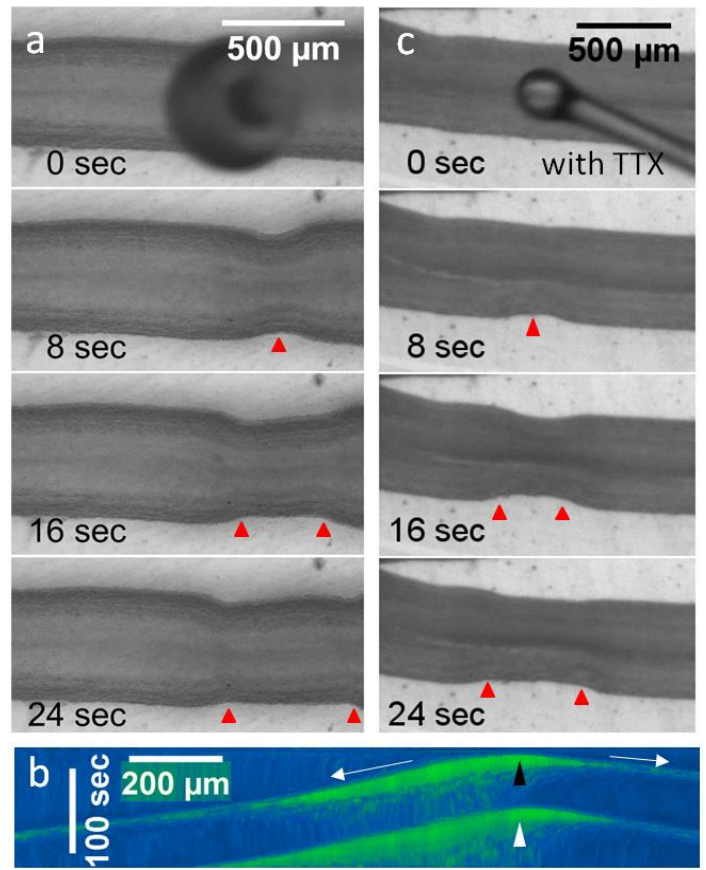

**Fig.3 : Pressure triggers propagating contractile waves. (a) Reaction to pinching of E9 gut (still shots from VideoS8). Red arrowheads point at the two outgoing contractile waves. (b) Motiligram derived from VideoS8 of the control gut. A pair of waves travels away (white arrows) from the point where the gut was pinched (black arrowhead). A second pair of waves nucleated at the same location ~30 sec later (white arrowhead). (c) Reaction to pinching of E9 gut after 30 min treatment with tetrodotoxin (1 µM, still shots from VideoS8).**

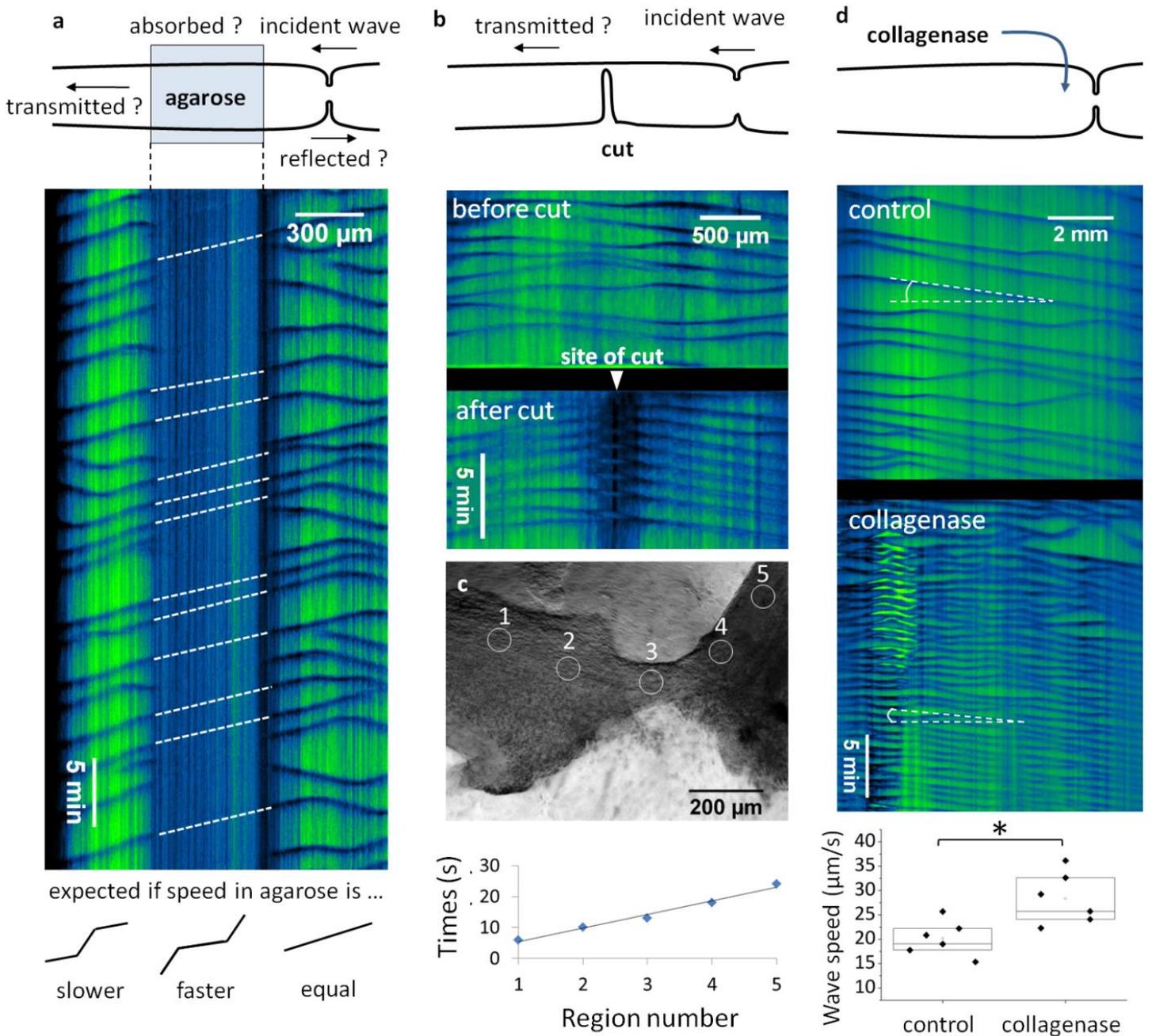

Fig.4 : Testing the role of mechanics in embryonic gut contractile wave propagation. (a) Agarose gel embedding experiment. Top: scheme of the experiment. Middle: representative motiligram, from VideoS9. Dashed white lines connect contractile waves propagating on either side of the gel. Bottom: explanatory scheme of expected motiligram depending on wave speeds in- and outside of the gel. (b) Effect of almost completely sectioning the gut on contractile wave propagation. Top: Scheme of the experiment. Bottom: Motiligram (E8 jejunum, from VideoS10) showing that the site where the cut is performed generates contractile waves in the ~10 min following the injury. (c) Almost completely sectioned midgut segment, and position of wave as a function of time in regions 1-5. (d) Effect of collagenase on motility (E8, midgut): motiligram and comparison of wave speed before/after collagenase application. Wave speed was deduced from the slope (white dashes) of the lines in the motiligrams. A star indicates a statistically significant difference (p<0.05, Mann-Whitney two-tailed test).

**Signaling by intercellular mechanical tension is not required for contractile wave propagation.** I have shown that the gut is sensitive to minute mechanical stimulation. It is therefore reasonable to hypothesize that the tensile forces generated by locally contracting cells may be felt by neighboring cells that, in turn, contract. This elementary propagation mechanism has recently been shown to direct contractile waves in the embryonic heart (8) and may also play a role in the embryonic gut. To test this hypothesis, I took inspiration from an elegant experiment designed by Johansson & Ljung to test the role of mechanics in adult vein contraction (10). To mechanically interfere with the contraction, I embedded a ~300 μm long gut segment in a drop of stiff agarose gel (Fig.4a and Materials & Methods). Gel embedding effectively reduced the contraction amplitude $A = (d_{rest} - d_{cstr})/d_{rest}$ ($d_{rest}$, $d_{cstr}$: resting and constricted diameter) from 7-8 % in agarose-free regions to <1% in the agarose embedded segment. The displacement of cells located within the gut was strongly reduced because of the new boundary condition imposed on the outer gut surface. I found that, in spite of the lack of gut wall displacement, peristalsis waves could travel unimpeded from one side of the agarose-embedded region to the other (VideoS9). In the motiligram of Fig.4a, wave propagation (slanted lines) in the non-embedded regions surrounding the agarose is clearly seen; I highlight propagation in the agarose with white dashed lines because the reduction of amplitude induced by the gel makes it is elsewise difficult to see. There was no discontinuity in the slope of the waves as they travelled through the agarose block (*n*=5), indicating that blocking the displacement of the outer wall did not affect wave propagation velocity (Fig.4a, bottom scheme).

In a second set of experiments, I disrupted the mechanical continuity of the gut by sectioning it almost completely (Materials & Methods). The two pieces were connected only by a flap of cells at the outer border of the gut (Fig.4b,c) placed perpendicularly (Fig.4c) to each other to minimize communication of tensile forces across the flap. I observed that in the first ~10 min after the cut was performed, the site of wounding became the source of numerous contractile waves traveling away from the cut (Fig.4b motiligram, VideoS10, *n*=5). After ~10 min the gut came back to a normal propagation pattern and waves traveling towards the cut in one segment could propagate through the flap of cells to the other gut segment (Fig.4c, *n*=5). The velocity of the wave remained constant as it traveled through the flap of cells (Fig.4c, bottom).

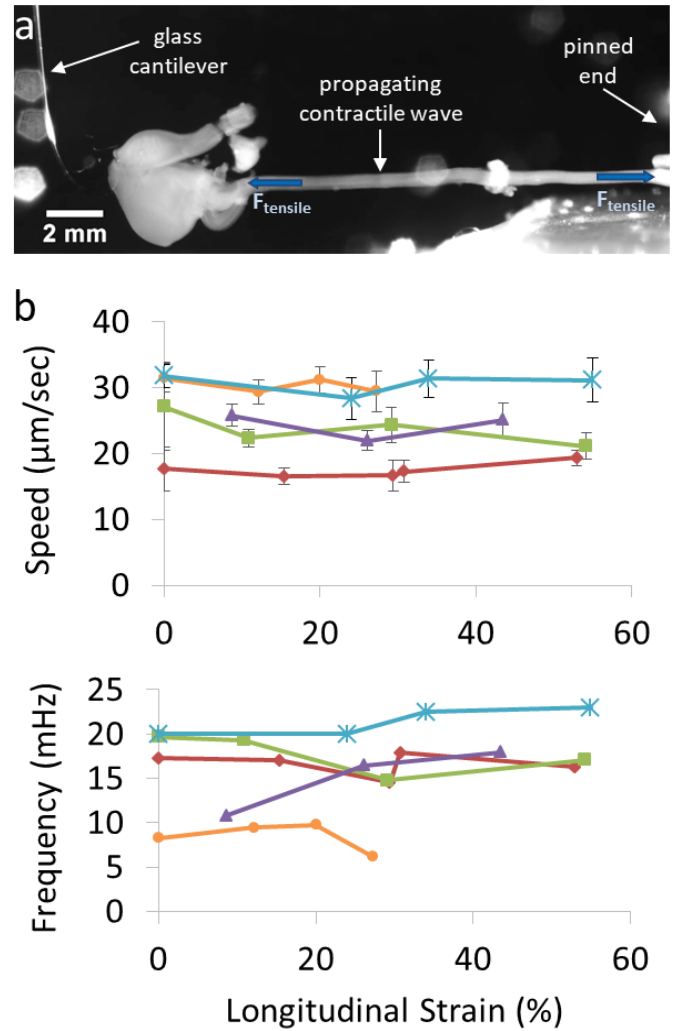

Figure 5 : Influence of longitudinal strain on contractile wave propagation and generation. (a) Experimental setup: the gut is pinned at one end and stretched at the other with a glass cantilever (12) (b) Speed and frequency of contractile waves as a function of applied longitudinal strain. Each color corresponds to a different gut (E9, *n*=4 and E8, *n*=1 in blue). Speed error bars represent the SD of n>4 waves. Frequencies were computed from 10 min periods spent at each strain value.

I next examined whether reducing the elastic modulus of the gut tissue using collagenase (cf Materials & Methods) would influence contractile wave propagation speed. Softening with collagenase has been shown to decrease propagation velocity in the embryonic heart; this effect was backed by a mechanical model of contractile wave propagation (8). I found that ~5 min after application, collagenase led to a near doubling (+80%) of the contraction frequency (Fig.4d). The average speed of the waves (Fig.4d bottom) before and after collagenase application was respectively 20.1 ± 1.8 μm/s (*n*=6) and 28.3 ± 2.7 μm/s (*n*=6)), i.e. collagenase induced a ~40% wave velocity increase.

I finally tested the influence of applying a longitudinal stress on contractile wave propagation. Changing the longitudinal pre-stress of muscle fibers could alter the propagation speed of contractile waves, much like increasing the tension of a violin string increases the velocity of deflection waves. Applying pressure has recently been shown to increase the frequency of smooth muscle contraction in the embryonic mouse lung (11). I applied tensile force by stretching the gut with a thin glass cantilever (12) (Fig.5a). Contractile wave frequency and speed of propagation in the jejunum were recorded for 10 min without strain, and then at 3-4 different, increasing strain values, for *n*=5 guts. I found that stretching in the strain range 0-50% (corresponding to a stress ~ 0-500 Pa) affected neither wave velocity nor wave generation frequency (Fig.5b).

## Discussion

We previously reported that a calcium channel blocker ($CoCl_2$) inhibits gut motility in the embryo, and that embryonic motility vanishes in calcium deprived medium (3); these observations are consistent with the idea that calcium activity underpins early embryonic gut contractility. The shape of the calcium spikes I measured (sharp up-rise followed by a slower decay, Fig.1d,e) is typical of calcium spikes in smooth muscle (13,14). A wide literature has been devoted to calcium waves and several of their properties give straightforward explanations of the dynamics of the early gut, in particular:

In some instances I observed contractile wave generation upon strain application, consistent with the mechanosensitive properties of the gut described above.

Embedding in gel (Fig.4a) or cutting (Fig.4b,c) did not prevent contractile wave propagation. In these experiments, I cannot exclude that some mechanical tension is still communicated from cell-to-cell, either by cells located in the gut interior (gel experiment) or through the cells located in the flap (cutting experiment). Wave velocity was however unaffected in either experiment, even though the mechanical properties of the medium in which the waves propagate were significantly modified. Modifying the longitudinal tensile pre-stress of muscle fibers did not affect longitudinal wave propagation speed or frequency (Fig.5). This suggests that mechanical cues do not play a central role in the propagation mechanism of early gut contractile waves. This conclusion is further supported by the fact that softening of the gut by collagenase led to an increase in wave velocity (Fig.4d), contrary to what is predicted by mechanical models of wave propagation in biological tissues (8)

1°) Early gut contractile waves propagate at constant speed, and with an undiminished amplitude (3). These are known properties of $Ca^{2+}$ waves (15). The contractile wave propagation speeds we measure are in the range 10-40 μm/s, i.e., similar to the speed of cytosolic calcium fertilization waves (16) and 2-3 orders of magnitude slower than embryonic heart (~20 mm/s (17,18)) or arterial smooth muscle (5-15 mm/s, (19)) calcium waves . The fact that the circumferential (along the fiber axis) propagation speed is higher than the longitudinal (perpendicular to the fiber axis) speed (Fig.1f) is consistent with previous measurements on adult gut (20).

2°) When two counter-propagating contractile waves meet along the digestive tract, they annihilate (VideoS1,S6). This is a direct consequence of the existence of a refractory period (21), i.e. the necessary time for CaATPase to reduce cytosolic $Ca^{2+}$ concentration back to its pre-excitatory level so that the tissue can become excitable again. Annihilation of calcium waves has been observed in cardiomyocytes rings (17) and I observe them here directly on gut explants (VideoS6).

3°) Wounding leads to the nucleation of numerous contractile waves at the injury site (Fig.4b). It is known that calcium waves are emitted after wounding of epithelia (22) or whole organisms (23). The calcium up-rise is probably due to influx from the extracellular medium due to disruption of membrane integrity after wounding (24,25), and has been shown to orchestrate wound healing (23,26).

4°) Calcium waves can be elicited in smooth muscle cell cultures by gentle mechanical stimulation (19,27). Bayliss & Starling (28) discovered more than a century ago that peristalsis movements in the adult (dog) intestine can be triggered by pinching or inserting bolus. The insertion of a bolus results in contraction of the muscle coat above the bolus, and its relaxation downward of the bolus. This behavior permits downward transport of the bolus and was coined the "law of the intestine". Here, I show that an essential part of this reflex – namely the ability to sense the bolus by the mechanical pressure it exerts on the gut wall - is present at the earliest stages of gut development in the embryo; it only involves muscle cells as ICCs are not yet differentiated at E6-E8 (29) and I found this reflex to be resistant to tetrodotoxin, a potent neuronal inhibitor. Pinching the embryonic gut led to the symmetric nucleation of two waves traveling away from the point where pressure was applied. Further development of a full, asymmetric reflex with ascending contraction and descending relaxation very likely requires coordinated input from enteric nerves (30). The stimulus I exert by pinching does not mimic pressure or distension within the lumen. I expect the latter to give rise to a similar reaction, because the gut was sensitive to minute stimulation (Video S8), and the reflex is unlikely to depend on the directionality of the mechanical excitation. It is reasonable to think that this reflex can be physiologically stimulated during embryogenesis, for example by the accumulation of amniotic fluid in the lumen. Note that contractile waves can also nucleate spontaneously (see all videos except VideoS8), i.e., their generation does not necessarily require mechanical stimulation.

Hao et al. (31) recently discovered spontaneous calcium waves in enteric neural crest cells in embryonic mouse gut at stages E12.5-E16.5. The waves were transmitted via purinergic P2 receptors and not gap junctions. Contractions of the gut could occur without ENCC $Ca^{2+}$ waves. The ENCC $Ca^{2+}$ waves subsided after stage E16.5, whereas contractions became more vigorous. The calcium activity I report here is located in the circular smooth muscle and contractions were always accompanied by a calcium transient unless the dye had bleached. As Hao et al. used a mouse strain in which only ENCCs express the calcium indicator, they could not observe the calcium activity in smooth muscle. I did not observe calcium transients in enteric neurons. The Fluo4-AM fluorescence signal tended to be saturated in enteric neurons (Fig.1b,c), which could have obscured readout of the calcium activity in these cells. However, in ganglia where the signal was not saturated (e.g. ganglia at 9:30, 10 and 11 o' clock in Video S3) or just partially saturated (e.g. ganglia at 8:30 o' clock in VideoS3), I did not observe any fluorescence spikes.

The fact that gap-junctional communication is required for contractile wave propagation/generation is consistent with previous work on $Ca^{2+}$ communication in smooth muscle cell cultures (14,32) and adult gut motility (33). Other small molecules that diffuse through gap-junctions, like ATP, may play a role in the propagation of the calcium wave (31,34). The blood vessels of a connexin-40 deficient mouse mutant have been shown to exhibit peristalsis-like activity (35). This suggests that a gut and a vein may differ, from the point of view of their dynamics, only by the amount of gap junctional coupling between adjacent circular smooth cells: high coupling results in synchronized contraction over long distances, i.e. vasomotion, whereas low coupling gives rise to local, propagative constrictions, i.e. peristalsis.

Chiou et al. (8) found that gap-junctions were not required for embryonic heartbeat propagation and that the heartbeat contraction velocity depended linearly on tissue stiffness. They concluded that early embryonic heart wave propagation is based on an active, local feedback to mechanical stretch. In contrast, my results indicate that, although the gut is mechanosensitive, the transmission of mechanical force from cell to cell during a contraction does not play a role in the propagation of the contractile wave. My experiments rather support a model in which propagation characteristics of the contraction are determined by the electrical conduction properties of the calcium wave across gap junctions. Johansson & Ljung (10) came to the same conclusion in their experiment on rat portal vein. In the adult, mechanical stimulation exerted by the bolus as it transits down the gut lumen has been shown (9) to be an integral part of its self-propulsion mechanism, by triggering sequential reflex contractions/relaxations. Development of this "neuromechanical loop" (9) likely requires enteric nerve activity, and is therefore not yet present at the embryonic stages I examine here.

Calcium waves propagating along a smooth muscle tube provide, at an embryonic stage, most of the essential dynamic features of peristalsis in the adult: the contractions obliterate the lumen, propagate, push luminal content and are mechanosensitive. The functional role of these calcium waves as a transient form of motility in the developing gastrointestinal tract remains to be elucidated: do these waves play a morphogenetic role? Do they transfer information to the already present, but quiescent enteric nervous system? These are some of the prominent questions that will require further investigation.

## Materials & Methods

**Specimen preparation & Ethics statement.** Fertilized chicken eggs (EARL Morizeau, France) were incubated from 6 to 9 days in a humidified 37.5°C chamber. The gut (hindgut to duodenum) was dissected out of the embryo and the mesentery was carefully removed. The experiments were conducted under European directive 2010/63/UE. The approval by an ethics committee is not required for research conducted on birds at embryonic stages.

**Calcium imaging.** Chicken guts were embedded in 4% agarose type VII gel and thin (50-200 μm) transverse slices were cut at the level of the jejunum with a razor blade. For calcium indicator loading, 50 μg of Fluo4-AM (Thermofisher, F14201) were dissolved in 10 μL DMSO; this solution was added to 2 mL of DMEM GlutaMAX$^{TM}$-I (Thermofisher, 4.5 g/L D-glucose, sodium pyruvate, $Ca^{2+}$ 1.8 mM, $Mg^{2+}$ 0.8 mM) containing 1% penicillin-streptomycin (PS), 25 mM HEPES and 0.01% Kolliphor EL (Sigma). Kolliphor (20) was essential for permeation of the dye in the smooth muscle cells. Up to 4 slices were placed in 1 mL of the loading solution for 30 min at RT. After washing, deesterification was performed in DMEM at 37.5°C in a 5% $CO_2$ incubator, for at least 30 min and at most 4 hours. Samples were shielded from light during the whole procedure. Imaging was performed on individual slices placed in a grazing layer of DMEM (Figure S1a) on an inverted Olympus confocal microscope, at magnification x40, laser excitation 488 nm, emission filter 512 nm, 1 Hz imaging frequency, 200 ms exposure-time, at 37.5°C. Images were collected from one confocal plane (no z-stack). Irreversible bleaching of the sample occurred after ~3 min imaging, i.e., contractions continued but the calcium concentration transients could not be observed anymore. This was not due to dye leakage as the transients of samples that had meanwhile been kept in the dark for up to 4 h could still be observed.

**Calcium fluorescence signal analysis.** Cells statically labeled by Fluo4-AM were used as fiducial markers. Coordinates $x(t)$ and $y(t)$ were retrieved with the Tracker ImageJ plugin (courtesy of O. Cardoso), yielding the displacement $d(t) = \sqrt{(x(t) - x(t=0))^2 + (y(t) - y(t=0))^2}$. An ImageJ macro computed the average fluorescence intensity in an elliptical ROI ~5-10 μm in size (~one cell) located at a fixed distance of no more than ~30 μm of the tracked point. The signal was normalized by $I_0$, defined as the minimal intensity recorded in this ROI in the image stack (video). Fluorescence changes due to displacements of the sample in and out of the confocal imaging plane were small and can be discerned from calcium spikes because the fluorescence intensity change they induce is smooth and overlaps with the displacement curve (unlike the calcium spikes in Fig.1d).

**Immunohistochemistry.** The slides were fixed 1h in 4% PFA, washed, blocked for 1 h in a 1% bovine serum albumin (BSA) and 0.1% triton in PBS solution, incubated overnight in anti-α smooth muscle actin (Abcam, 5694, dilution 1:1000) and anti βIII-tubulin antibody (Abcam 14545, dilution 1:1000). After washing, CY3- and A488-conjugated secondary antibodies (ThermoFisher, dilution 1:400 in PBS) were applied for 2 h at RT. The slides were washed and imaged at x40 with the confocal microscope. The images were rotated to have the same orientation as on the calcium activity recordings. I used the shape of the epithelium and the localization of enteric ganglia as guides to find the correct rotation angle.

**Whole-gut motility monitoring setup and motiligram generation.** Guts were placed on a porous alumina membrane (Anodisc™, Fig.S1b) resting on the edge of a tissue culture dish (Ø~20 mm) filled with DMEM + 25 mM HEPES as previously described (3), in a humidified Petri-dish at 37.5°C. Motiligrams (Fig.2-4) were generated from the image stacks by tracing a rectangular ROI along the edge of a midgut segment and by using the ImageJ function "Reslice", see (3) for details. Drugs and chemicals were added directly to the culture medium underneath the Anodisc.

**Drugs & Chemicals.** Tetrodotoxin 1 μM (Abcam). Collagenase-dispase 0.33 mg/mL (Roche). The gut can be dissociated by pipetting after 5 min at 37°C at this concentration, indicating that softening occurs within the first minutes after collagenase application. Enoxolone (18β-Glycyrrhetinic acid, Sigma-Aldrich) 33 μM (DMSO 1‰) and 100 μM (DMSO 3‰). DMSO alone does not affect gut motility at these concentrations. Heptanol 100 μM and 500 μM (Sigma-Aldrich). A 10 mM heptanol concentrate in PBS solution was prepared, vigorously agitated to homogenize the two-phase solution (milky white) and diluted in the sample medium.

**Mechanical Manipulations of the Embryonic Gut.** To block displacement of the gut wall, a drop of agarose type VII 4% w/w was placed with a thin syringe needle on a portion of midgut lying on an Anodisc membrane and allowed to solidify at RT for ~5 min. Exact position of the agarose-embedded segment could be visualized by phase-contrast imaging of the sample prior to time-lapse imaging. Precision cuts of the GI tract were realized with a surgical micro-scissor (Euronexia). Longitudinal stress/strain was applied to the gut as previously described (12), in DMEM at 37°C.

## Acknowledgments

I thank Vincent Fleury for proofreading the manuscript, Nicolas Dacher for help performing the experiment in Fig.5 and Max Piffoux for suggesting to test the reversibility of enoxolone.

## Funding

This research was funded by a CNRS / INSIS Starting Grant "Jeune Chercheur" and by the Labex "Who Am I ?" (ANR-11-LABX-0071).

## Data Availability Statement

All relevant data are within the paper and its Supporting Information files. Higher resolution videos are available from the author upon request.

## Competing Interests

I have no competing interests.